\documentclass[10pt]{article}
\usepackage{amsmath}
\usepackage{amssymb}
\usepackage{graphicx}
\usepackage{subfigure}
\usepackage{cite}
\usepackage{changes}

\usepackage[labelfont=bf,labelsep=period,justification=raggedright]{caption}
\makeatletter
\renewcommand{\@biblabel}[1]{\quad#1.}
\makeatother
\date{}
\begin{document}

\begin{flushleft}

{\Large
\textbf{A closed parameterization of DNA--damage by charged particles, as a function of energy --- A geometrical approach}
}

Frank Van den Heuvel PhD$^{\ast1,2}$, 
\\
\bf{1} CRUK/MRC  Oxford Institute for Radiation Oncology, Department of Oncology, University of Oxford, Oxford, UK \\
\bf{2} Laboratory for experimental radiotherapy, Department of Oncology, University of Leuven, Leuven, Belgium\\
$\ast$ E-mail: frank.vandenheuvel@oncology.ox.ac.uk
\end{flushleft}
\section*{Abstract}
{\em Purpose:} To present a closed
formalism calculating  charged particle radiation damage induced
in DNA. The formalism is valid for all types of charged particles and
due to its closed nature is suited to provide fast conversion of
dose to DNA-damage. 
\newline
{\em Methods:}
The induction of double strand breaks in  DNA--strings residing  in irradiated cells
is quantified using a single particle model.
This leads to a proposal to use the cumulative Cauchy distribution to express the mix 
of high and low LET type damage probability generated by a single particle.
A microscopic phenomenological Monte Carlo code is used to fit the parameters of the model as a function of kinetic energy
related to the damage to a DNA molecule embedded
in a cell. The model is applied for four
particles: electrons,
protons, alpha--particles, and carbon ions. 
A geometric interpretation of this observation using the impact ionization mean free path as a quantifier, allows extension of the model
to very low energies.
\newline
{\em Results:} 
The mathematical expression describes the model adequately using a chi--square test ($\chi^2/NDF < 1$).
This applies to all particle types with
an almost perfect fit for protons, while the other particles seem to
result in some discrepancies at very low energies.
The implementation calculating a strict version of the RBE based on
complex damage alone is corroborated by experimental data from the measured RBE. 
The geometric interpretation generates a unique dimensionless parameter $k$ for each type of charged particle. In addition, it predicts a distribution of 
DNA damage which is different from the current models. 
\section*{Introduction}
The biological effect of ionizing radiation on human cells is believed
to be related to the generation of damage in the
DNA--molecule located in the cell's nucleus\cite{Hall}. The physical mechanism
is the ionization of the DNA macro molecule, generating lesions in the molecular structure,
either by direct ionization 
or by the generation of radicals in the vicinity of the DNA which then
indirectly
damage it. These events (direct or indirect) can create several types of damage
to the DNA by combining a number of lesions into a cluster, which can
only happen if they occur in close proximity (typically within one turn of
the DNA--helix).
The most prevalent of these damage types are base damage (2 lesions), followed by single strand
breaks (SSB) (3 lesions) , double strand breaks (DSB) (4 lesions), and locally multiple damage sites
(LMDS). The latter are clusters of different types of damage occurring close to
each other.  
It is shown that base as well as SSB damage is not likely to be a
deciding factor in the destruction of cells, due to the efficient repair
mechanisms which exist in the cell\cite{Caldecott2008}. The combination of double
strand breaks and LMDS's is likely to be the root cause for cell kill\cite{Ward1985}.
\par
To quantify the amount of ionizing interactions in a medium, the physical notion of dose can be used.
Dose is defined as the amount of energy deposited in a medium per unit mass and is expressed in Joule(J) per kg or Gray (Gy).
In the case of dose deposition by charged particles 
the Bethe--formalism is used. This describes ionization events in a medium in terms of
energy loss of the charged particles in inelastic collisions with the
electrons of the medium, through the notion of mass stopping power
($dE/\rho dx$). In his seminal work already in 1930, Bethe showed that there is an intimate relationship between
stopping power on the one hand, and energy (i.e. speed), charge, and the medium in which the interaction takes place on the other hand\cite{Bethe30}. 
A further extension taking into account the possibility of the charged particle picking up electrons, thereby changing the stopping power was introduced by Barkas\cite{Barkas1963}, using the concept of an effective charge. 
\par In radiation biology, 
linear energy
transfer (LET) is used rather than stopping power. 
LET is identical to stopping power with the energy delivered to $\delta$--rays (i.e. highly energetic knock on electrons)
subtracted. 
This quantity is called restricted stopping power. As such, LET is a  measure for the density of ionization taking
place along the track of a charged particle through a medium.
Due to its close relationship with stopping power, it follows that there is a close relationship between LET and the kinetic
energy of the depositing particle.  
From
observation a dearth of DSB's and LMDS's was shown to be related to low LET irradiations,
while an increased number of both  for the same dose is seen  high in hight LET irradiations\cite{Hall}.
Brenner and Ward\cite{doi:10.1080/09553009214551591} argued that
DSB and LMDS damage was related to multiple interactions by single particles, rather than the combination of single strand breaks generated by
single particles. In the field of microdosimetry, this is taken a step further by defining the notion of lineal
energy which introduces the amount of energy deposited along lines confined in a convex geometric shape
 with a given distribution of cord lengths estimating the energy deposited in
various shapes, which can be used for measurement (i.e. spheres, cylinders). 
\par
Extending this, it is natural to propose a model where distance between ionizations  along these lines 
plays a significant role in the generation of DNA--damage.
A full listing and treatment of these quantities can be found in the ICRU reports 16, 19, and 36
\cite{icru16,icru19,icru36}.
\par
To describe the damage impact of charged particles on the
DNA--structure, the science community has taken its recourse to using
Monte Carlo simulations
to quantify the damage introduced\cite{Nikjoo1997,Stewart2011}.
A more fundamental
analytical approach is currently lacking, due to
the underlying complexity of the DNA molecule, and the paucity of the available
experimental data. The data which is available is mainly provided in terms of
relative biological effective dose (RBE), a quantity combining physical,
spectral, chemical, and biological factors,
all of which hamper ab--initio
calculations. 
\par
Monte Carlo calculations are able to predict the induction of simple or complex
damage as well as induction of single and double strand breaks in DNA--molecules. These
findings are interpreted using the Bethe--Barkas formalism in terms of LET and
show that high LET particles indeed introduce more complex damage. 
\par
In this paper we develop a parameterization using a simple geometrical model,
that describes the behavior as calculated by the Monte Carlo codes.
We also show that this formalism describes the current knowledge well.
\section*{Methods and Materials}
\subsection*{Theory}
\begin{figure}[t]
\centering
\includegraphics[width=0.6\columnwidth]{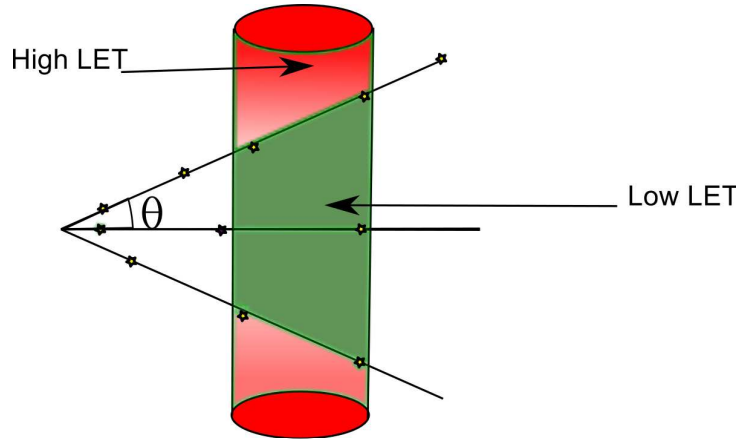}
\caption{\label{cylinder_model} A schematic model of a source of charged
particles with a given mean free path length (i.e. a
given energy), which is comparable with the diameter of the sensitive
cell volume.
As the angle ($\theta$) of the particle's path with respect to the normal to
the axis of the structure increases the chance that more than a single
event will occur in the volume. This implies that if the angle ($\theta$) is larger than the one for which
the projection of the average length between interactions equals the diameter, 
more high LET events will be registered.}
\end{figure}

We use the single charged particle
model as proposed by Brenner and Ward, 
distinguishing three
types of interaction results: Low LET, high LET, and intermediate LET
mode. The specifics of each mode are explained below.
\begin{enumerate}
\item Low LET: A single  particle is generally not able to generate lesions close enough together 
to induce double strand breaks at each interaction. 
It is clear that DSB's can be generated but in a limited fashion and that we use the word lesion in liberal fashion to indicate an interactive event which has damage as a consequence.
\item High LET: The particle has the possibility to generate multiple lesions 
irrespective of any geometrical considerations. We implicitly assume that the double strand break damage is the result of multiple interactions by one particle. How exactly this damage is
introduced (direct or indirect) is outside the scope of this article. An implicit assumption
however is that ionizing events need to be geometrically close to the
DNA structure.
\item Intermediate: In given geometric circumstances it is possible for the
charged particle to generate DSB--damage, in a high--LET manner, depending on the angle
under which the
particle hits the sensitive volume (Fig. 1). 
\end{enumerate}
As a surrogate to categorize the charged particle in one of the types
defined above, we use the mean path length between ionizing interactions
in a medium consistent with the atomic make up of a DNA--molecule for
the type of particle under consideration. In the remainder, we denote
this with $\lambda(E)$, where $E$ is the kinetic energy of the particle.
If $\lambda(E)$ is large relative to the sensitive volume, then the lesions on average are too far apart
and only damage types related to a few lesions can occur (i.e. SSB and base damage).
Charged particles with such energies will be part of the first
category. If on the other hand $\lambda(E)$ is small then the probability
of lesions creating more complex clusters of damage close together will be higher. 
Charged
particles with this property will be in the high LET category. Finally,
charged particles with intermediate distances between ionizing events
have the capability of generating DSB and LMDS damage depending on other
factors than $\lambda(E)$ alone. In this model we use the geometric
direction of the path of the charged particle as a parameter.
In Figure \ref{cylinder_model} a schematic model of this approach
is shown. This implies that only a limited amount of directions are
available to
contribute to the amount of complex damage in the manner as outlined
for the high--LET type interactions. This happens when for a particle of
a given energy
the quantity $\lambda(E)$ is slightly larger than the maximal distance
between two DNA--damage  lesions to be considered as being in the same cluster
(usually about 10 base pairs (bp)). Due to the finite
thickness of the sensitive  volume it is possible to behave in a high--LET
fashion depending on the angle with which the particle's path crosses
the volume. This occurs when the projection of the path is smaller than
the previously determined maximum.
\par
\par
\subsection*{Equivalence Principle}
In the case of irradiation with charged particles all directions
of the particle's paths are possible as are all rotational positions of the
DNA--structure.
A particle that interacts (i.e. that creates a lesion) at the surface of a given 
sensitive volume has limited possibilities to interact again given that on average,
a specific distance (which depends on the particle energy) has to be
travelled before it interacts again. The next interaction's position
is then limited by the constraints outlined above if it is to fall within
the sensitive volume. 
This first interaction can happen anywhere along the volume, but the
constraints are relative to the position of that point. This implies
that we can invoke an equivalence principle and reduce the problem to
that of an isotropic point source positioned at the surface of a sensitive volume.
\subsection*{Mathematical expression of the equivalence principle}
We need to calculate what fraction of the paths starting in the given
point can interact with the sensitive volume given
the fact that there is a length within which this is not likely,
provided by $\lambda(E)$, and that there is a maximal distance ($H$)
that disqualifies the
generated lesion to be registered in the same cluster.
We have reduced this problem to that of the distribution of  
projections of a point source on a line--piece, the  solution to which  is
known as the Cauchy--distribution\cite{cauchy:1853},  and is described by the Lorenz function $f(x)$ with $x \in \Re$ expressed as follows:
\begin{equation}
f(x)~=~\frac{1}{1+ (\frac{x}{r})^2}
\end{equation}
\begin{figure}[h]
\centering
\includegraphics[width=0.45\columnwidth]{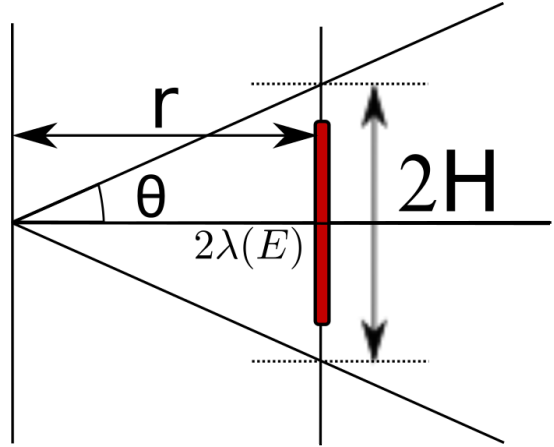}
\caption{\label{geom_model} 
The abstracted version of Figure \ref{cylinder_model} describes the distribution
of horizontal distances at which a line segment tilted at a random angle
$\theta$ cuts the x--axis. Only particles with large angles 
contribute to double strand break
events by combining damage generated by a single particle. The red line
indicates the ``forbidden'' area as (on average) this  distance between the ionization
events is observed.}
\end{figure}

From Figure \ref{geom_model}, it follows that the contribution $P$ for a given energy of charged particles to high
LET events is proportional to:
\begin{equation}
P ~ \sim ~2\int_{H-\lambda(E)}^H \frac{1}{1+ (\frac{x}{r})^2} dx
\end{equation}
Performing the calculation, we obtain:
\begin{align}
P ~& \sim ~\frac{2}{\pi}\lbrack \tan^{-1}(\frac{x}{r}) \biggm|_{H-\lambda(E)}^H\rbrack + C \\
~& \sim ~\frac{2}{\pi}\lbrack\tan^{-1}(\frac{H}{r}) - \tan^{-1}(\frac{H-\lambda(E)}{r})\rbrack + C\\
~& \sim ~\frac{2}{\pi}\lbrack\tan^{-1}(\frac{\lambda(E)-H}{r}) + \tan^{-1}(\frac{H}{r})\rbrack + C
\end{align}
This implies that the amount of DSB--damage for a given dose and given energy of the charged particle is governed by the following expression.
\begin{equation}\label{Cauchy}
F_{cd}(E) ~= ~(a-b)\frac{2}{\pi}\lbrack\tan^{-1}(\frac{\lambda(E)-H}{r})\rbrack + b
\end{equation}
The change from a low to a high LET regimen occurs over a small energy
interval. In such a small interval the average distance dependence on the
energy of the particle can be approximated with a linear function. 
Therefore, we expect the
energy dependence of the contribution of complex damage to follow the
same form as in Equation \ref{Cauchy} yielding the following expression,
with H=$\lambda(E_0)$ and r=$\lambda(\Gamma/2)$, $E_0$ being the energy, where the change in DSB is maximized and $\Gamma$ a measure for the width of the slope (i.e. the full width at half maximum in differential energy space).
\begin{equation}
\label{response}
F_{cd}(E) ~=~ (a-b)\frac{2}{\pi}\lbrack\tan^{-1}(\frac{E-E_0}{\Gamma/2})\rbrack + b
\end{equation}
With the parameters $a$ and $b$ related to the levels 1 and 2 as outlined
above. From boundary conditions we find that at very large energies
(i.e. $E >> E_0$) the expression is reduced to minimal number of double
interactions ($D_{min}$) which is equal to $a$. The value of $b$
is related to the maximal number of double interactions ($D_{max}$)
as follows:
\begin{equation}
D_{max} = D_{min} + (1 - \tan^{-1}(-\frac{E_0}{\Gamma/2})) b .
\end{equation}
The formalism using energy alone allows us to forego specific assumptions
regarding the dimensions of the DNA--molecule. Furthermore, it also
allows us to apply this technique to particles where the values for $\lambda(E)$ are less well known.
In addition, it allows us to test this formalism using experimentally available data which is available as a function of energy.
\subsection*{ Validation using Monte Carlo Simulations}
The use of microdosimetric calculations has provided important insight into the
mechanisms and effects of radiation deposition. In the past, Monte Carlo simulations
of charged particle deposition by various modalities were used to quantify and
typify the kinds of damage introduced by the different modalities\cite{Nikjoo1997}.
\par
The  Monte Carlo Damage Simulation code (MCDS) developed by Semenenko and Stewart,
generates spatial maps of the damaged nucleotides forming many types of clustered DNA lesion,
including single-strand breaks (SSB), double strand breaks (DSB), and
individual or clustered base damages\cite{Stewart2011}.
This approach has been shown to yield a linear relationship of the number of generated DSB's up to a high dosage. It follows 
that this parameterization also provides the possibility to link dose to damage.
In this paper, MCDS version 3.0 was used with the 
parameters described below.
The DNA length  which was chosen to be 1Gbp (Giga base pairs) and a nucleus
diameter of 5$\mathrm{\mu m}$. 
In the MCDS software, the geometry of the DNA--molecule is not an explicit parameter. 
Here four parameters are used: 1) the DNA--segment length $n_{seg}$, which is an {\em ad hoc} parameter expressed as 
base pairs $Gy^{-1}cell^{-1}$, 2) the number of strand breaks generated $\sigma_{Sb}$, 3) the number of base pair damages 
generated $\sigma_{Bb}$ by defining $f=\sigma_{Bb}/\sigma_{Sb}$, and 4)
a parameter $N_{min}$ (bp) describing the minimal separation for damage
to be apart not to be counted as being in the same cluster. The values
of these parameters is determined on the basis of other simulations
and measurements. For a more in--depth treatment of these parameters we
refer to the work by Semenenko and Stewart\cite{ISI:000237044600004}.
Variable input parameters MCDS were; the modality
(i.e. energy depositing particle (electron, proton,\ldots)), the energy
(in MeV), and the oxygen concentration in \%.
In the implementation described here we 
chose to omit any oxygen enhancement
as this could be a confounding factor and is the subject of another study.
In this study it was found that oxygen only changed the amount of damage in the 
low LET regimen, leaving the formalism unchanged (data not shown). 
Therefore, a concentration of 0\% oxygen was used.
For every particle type  at the relevant kinetic energies,
all complex damage
was noted per Gy, per cell and per kinetic energy.
\subsection*{Fitting procedure}
The ultimate goal was to fit the complex damage function to the data obtained by the Monte Carlo simulation. 
The parameters that need fitting are the energy position ($E_0$) the width
of the underlying Cauchy distribution ($\Gamma$) and the  parameters $a$
and $b$.
If a regular fit (i.e. all parameters fit at the same time) is performed we see strong  co--variances between the parameters.
To come to meaningful results we opted to perform a two step procedure:
First, we eliminate the parameters $a$ and
$b$ by fitting the differential, thereby  reducing expression \ref{Cauchy}
to the Lorenz function. 
\begin{equation}\label{BW}
\frac{dF_{cd}}{dE}=\frac{\Gamma^2/4}{(E-E_0)^2 + \Gamma^2/4}
\end{equation}
This is also mathematically equivalent to the fit of a Breit--Wigner resonance in high energy physics\cite{Breit59}.
In a second fit--procedure, the remaining variables $a$ and $b$ are fit using the cumulative Cauchy function. 
The fitting procedures were performed in the 
gnuplot\footnote{http://www.gnuplot.info}--software using a Levenberg--Marquardt minimization routine.
\section*{Results}
\begin{figure}[h]
\centering
\subfigure[\label{electron_BW_fit} Electrons]{\includegraphics[width = 0.45\columnwidth]{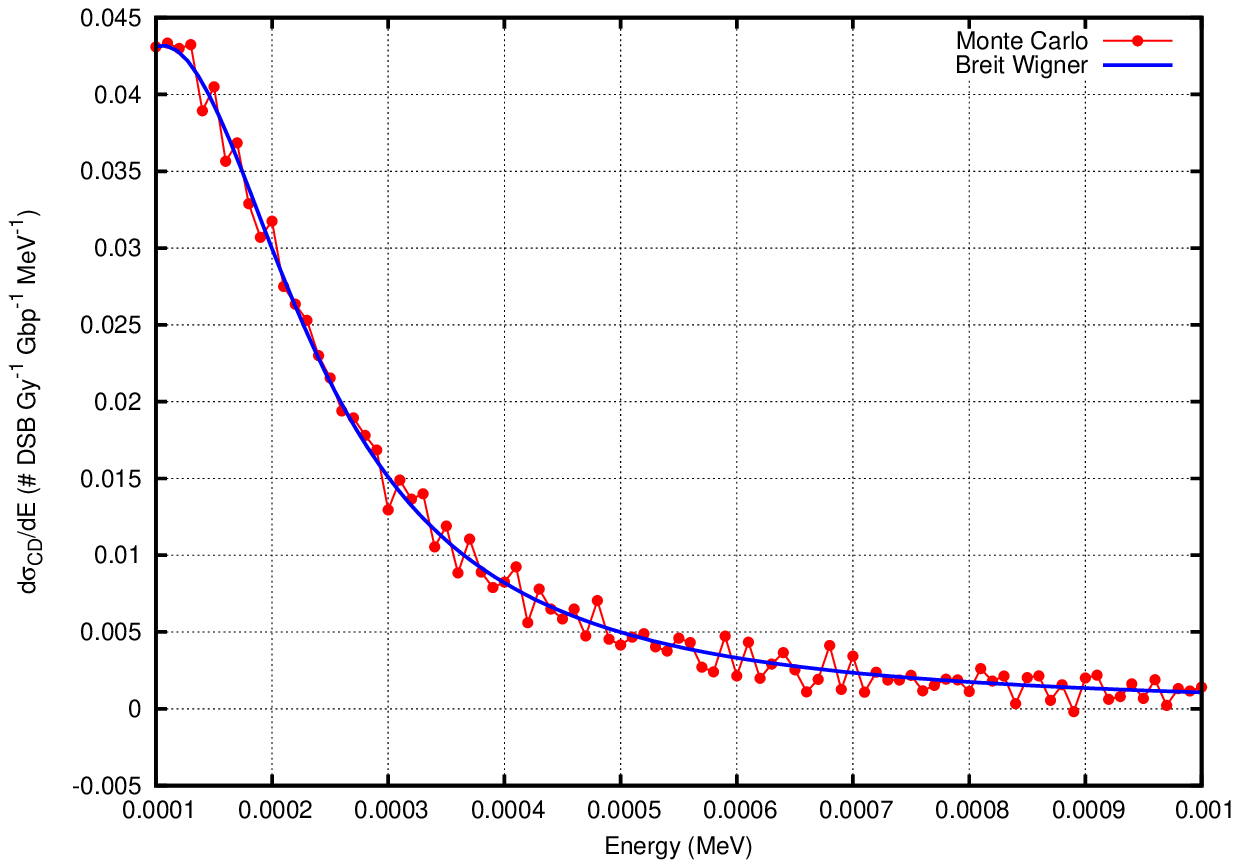}}
\subfigure[\label{proton_BW_fit} Protons]{\includegraphics[width = 0.45\columnwidth]{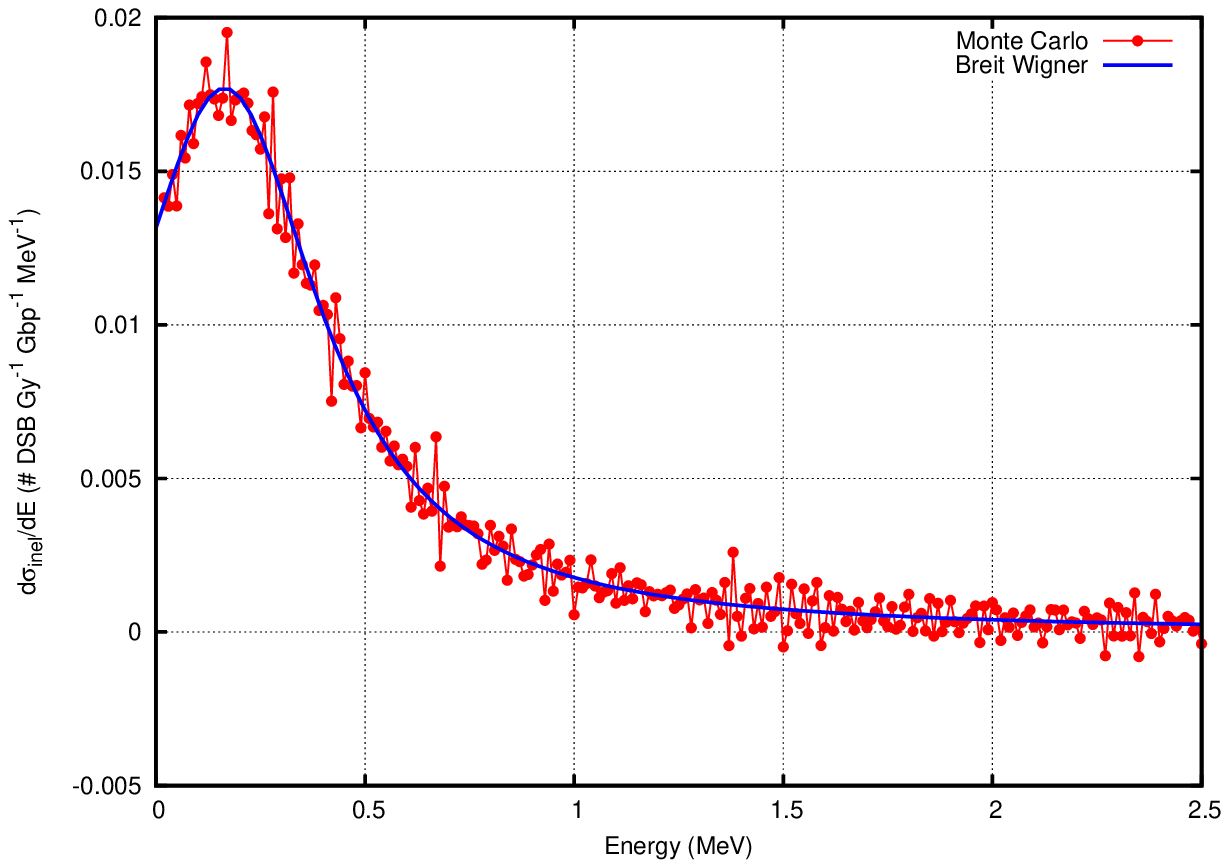}}
\subfigure[\label{alpha_BW_fit} $\alpha$--particles]{\includegraphics[width = 0.45\columnwidth]{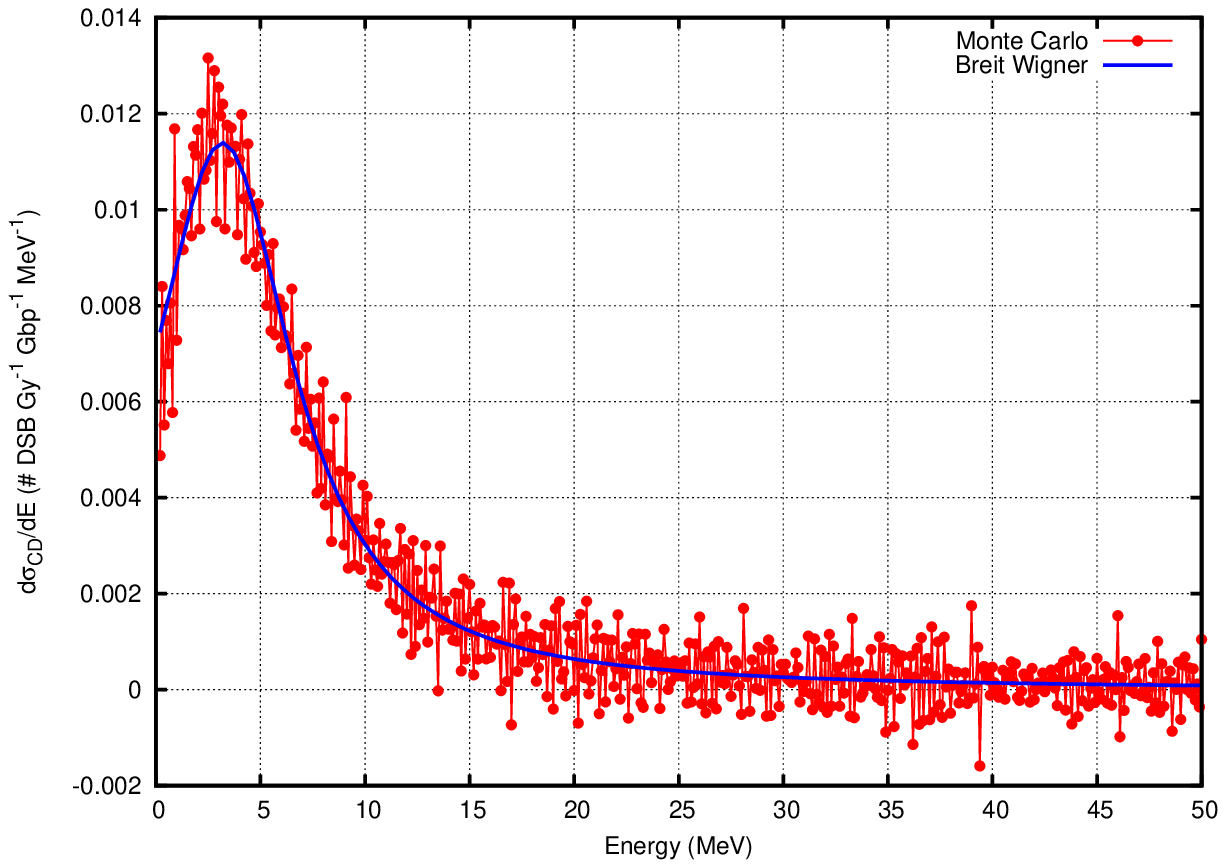}}
\subfigure[\label{carbon_BW_fit} Carbon Ions]{\includegraphics[width = 0.45\columnwidth]{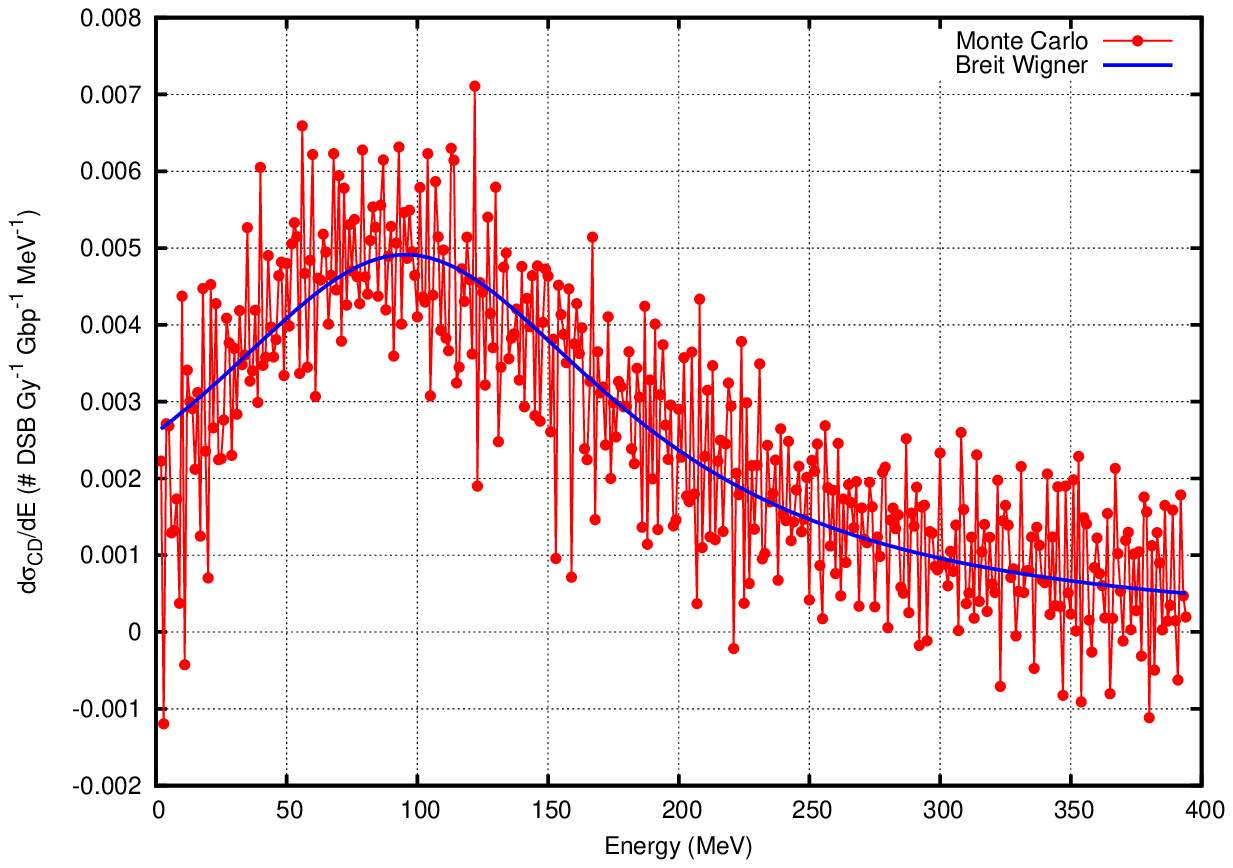}}
\caption{\label{BW-fit} Fitting the Cauchy expression to the energy differential probability of generating DSB's denoted $\frac{d\sigma}{dE}$}
\end{figure}

In Figure \ref{BW-fit},  the Lorenz expression as outlined in Equation \ref{Cauchy} together with a normalization factor,
is used to fit the energy differential probability for the generation of DSBs. The fit is
performed to minimize the
$\chi^2$--value. In all cases, the resulting  $\chi^2/NDF$ (NDF = Number
of Degrees of Freedom) are lower than 1. The values of the parameters
are provided in Table \ref{BW_table}.
\begin{table*}[ht]
\centering
\begin{tabular}{| l | c | c | c | c |}
\hline
Particle&$\Gamma$&$E_0$&$a$&$b$\\\hline
$e^-$&(2.854$\pm$0.051)$10^{-04}$ MeV&(1.05736$\pm$0.036)$10^{-04}$ MeV&2.9061&21.460\\\hline
$p^+$&0.5575$\pm$0.0094 MeV&0.1642 $\pm$ 0.0037 MeV&2.89068&21.4273\\\hline
$\alpha^{++}$&8.20$\pm$0.17 MeV&3.1850$\pm$0.056 MeV&3.0856&20.7933\\\hline
C$^{6+}$&201.7$\pm$8.4 MeV&95.4$\pm$2.5 MeV&3.01459&21.8489\\\hline

\end{tabular}
\caption{\label{BW_table} The different values for $\Gamma$ and $E_0$
as defined by Equation \ref{BW} and obtained from a fitting procedure
together with the asymptotic standard error of the fitted parameter. All
fits exhibited minimal values of $\chi^2/NDF$ (NDF = Number of Degrees
of Freedom). The columns $a$ and $b$ are the parameters indicating the
levels of DSB at low, resp. high LET. Note that even in low LET the
number of DSB's is not zero as complex damage can occur due to the combination
of simple damage events.}
\end{table*}
All fits are completely satisfactory at energies higher than $E_0$. On the lower energy side 
some discrepancies can be observed depending on the incoming
particles, particularly in the case of electrons and carbon ions. We refer the reader to the discussion section. For protons we see a satisfactory fit over the full
energy range.
\par
\begin{figure}[h]
\centering
\subfigure[\label{electrons_DSB_fit} Electrons]{\includegraphics[width = 0.45\columnwidth]{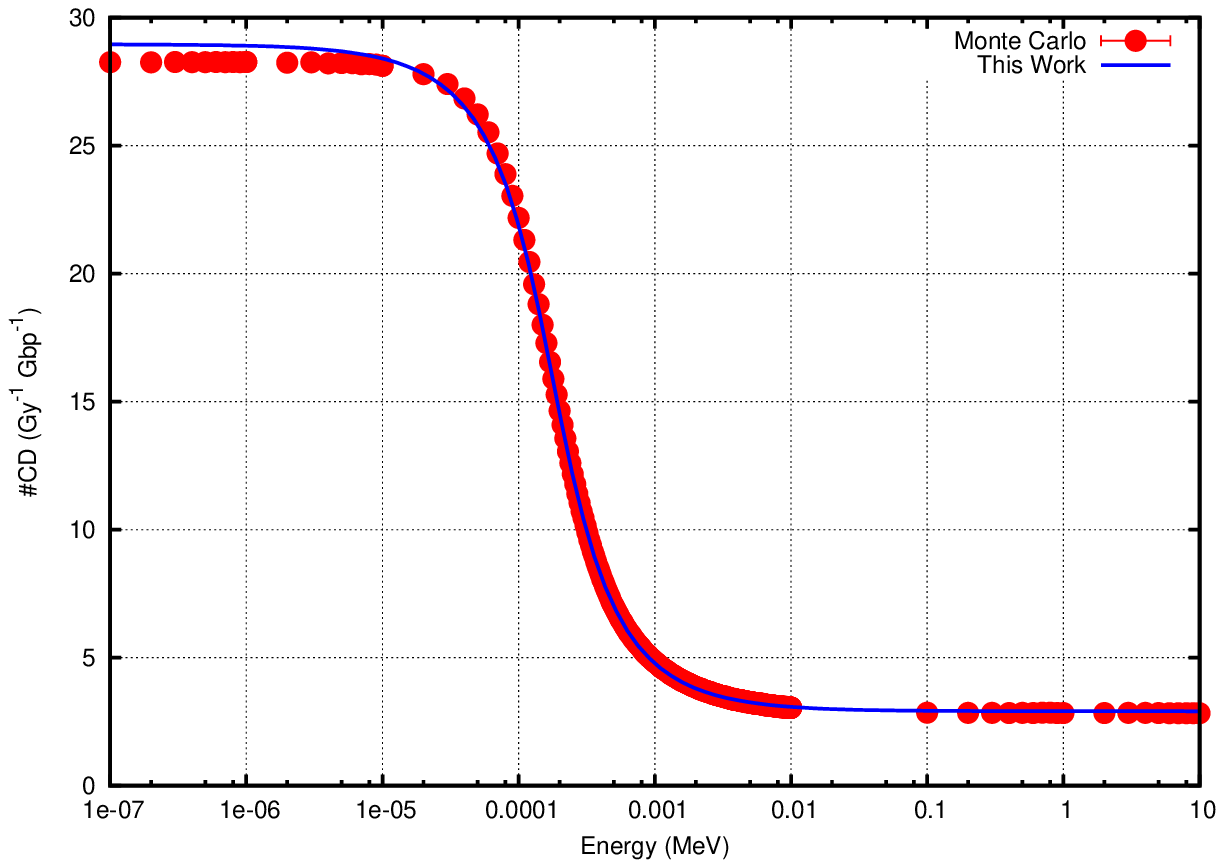}}
\subfigure[\label{proton_DSB_fit} Protons]{\includegraphics[width = 0.45\columnwidth]{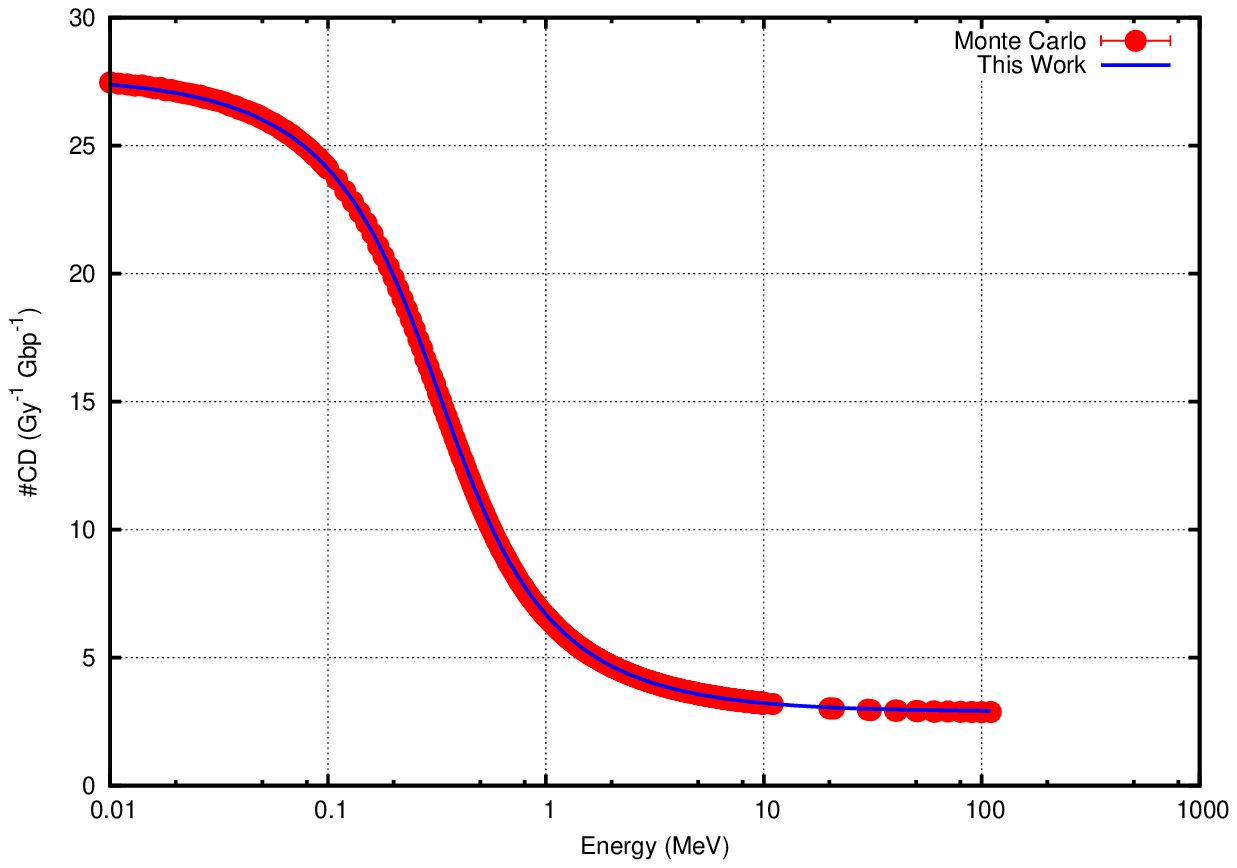}}
\subfigure[\label{alpha_DSB_fit} $\alpha$--particles]{\includegraphics[width = 0.45\columnwidth]{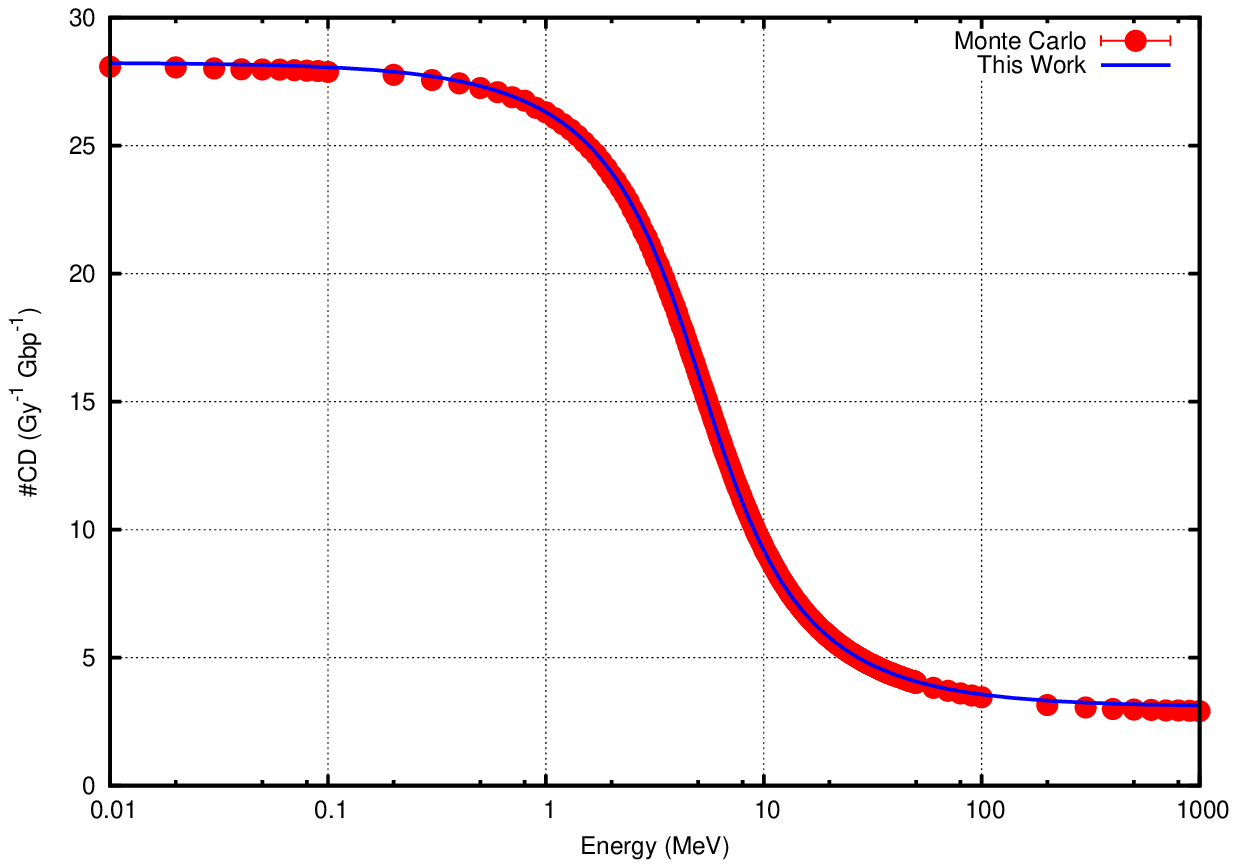}}
\subfigure[\label{carbon_DSB_fit} Carbon Ions]{\includegraphics[width = 0.45\columnwidth]{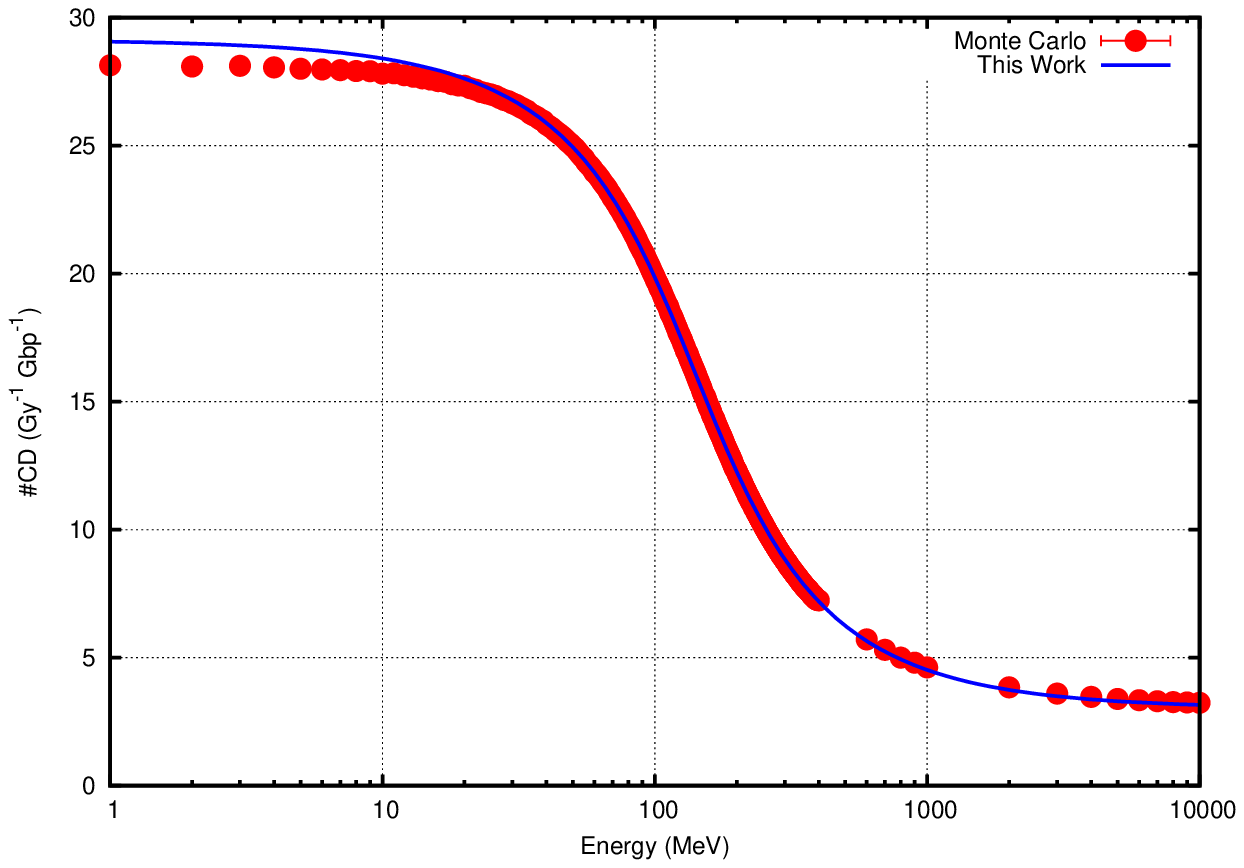}}
\caption{\label{BW-fit_cul} The prediction of the number of double strand breaks or more complex damage as a function of energy for 4 relevant charged particles. This provides the number of Double Strand breaks (DSB) per Gy, Gbp and per cell.  The prediction for protons and alpha particles is almost perfect. For electrons and carbon ions some discrepancies exist at lower energies.}
\end{figure}

 
Figure \ref{BW-fit_cul} shows the final results with all parameters fit. Again, all fits have $\chi^2$--values commensurate with a positive goodness of fit. 
The final values and the standard errors for the fitted parameters are listed in Table \ref{BW_table}.  
Note, that the noise in the differential curves increases as the particles become heavier. 
The random-seeming errors in the estimates of the derivative arise in part from the Monte Carlo estimates of the mean number of DSB per Gy per Gbp and from numerical instabilities associated with the calculation of the derivative using finite difference methods.
\subsection*{Geometric approach}
Now is the time to investigate the geometric interpretation further. 
To quantify the function $\lambda(E)$ we can use the inelastic mean free path as measure (IMFP).
Values for IMFP for electrons are
well known in the literature, not in the least as they are important in
solid state physics and electron microscopy. They can be found in freely
available databases for a variety of elements and compounds, even for
organic molecules like DNA\cite{NIST2010}. Proton values can be
found in a publication by Zhen--Yu and colleagues\cite{TANZhen-Yu:113403}.
For heavier particles such as $\alpha$--particles and carbon--ions, the data is more difficult to find.
We therefore opt not to use the data for these particles and restrict ourselves to electrons and protons in this further treatment.
\par
In all current microdosimetric codes, the Bethe formalism is used which
is valid for higher energies
(i.e. above 500eV for electrons). This implies that changes in IMFP, denoted by $\lambda$,
which impact the damage calculated using these codes, also reflect the
limitations of the
Bethe formalism.
From the theory the following expression is used:
\begin{equation}\label{first_appr}
\lambda(E) =  \frac{E}{A \log (E/E_0) + B}
\end{equation}
\begin{table}[h]
\centering
\begin{tabular} { | l | c | c |}
\hline
Particle&A&B\\\hline
Electrons&69.200 eV/nm&-153.94 eV/nm\\\hline
Protons&115.231 keV/nm&-301.45 keV/nm\\\hline
\end{tabular}
\caption{\label{ziaja_table} Parameters obtained by fitting Eq. \ref{first_appr} to data obtained from NIST (electrons) and Zhen Yu et al. (protons)}
\end{table}
\par
In this work the parameters $H$ and $r$ have thus far not been linked
to any physical property but were fit. An interesting proposition could be to link
these to dimensions of the target structure. Indeed, the choice of
a cylinder as a geometric representation is not an accident. It is
natural to
use the diameter of a DNA--molecule as a measure of the cylinder's
diameter. The length of the cylinder is then related to the maximal
distance we allow to classify two damage events, being part of the
same cluster of complex damage. Both values can readily be found in the
literature and text books\cite{sinden1994dna}.
For the most prevalent form  of cellular DNA (B--DNA), the values are
$3.4$nm (i.e. the height of a spiral of 10 base pairs), and
$2.37$nm as the diameter. We now define a dimensionless quantity $k$
which is specific to the type of charged particle used.
It is clear that this parameter acts as a scaling parameter but also
depends on the ratio of both fixed parameters.
Equation \ref{Cauchy} now reads as follows:
\begin{equation}\label{Cauchy2}
F_d(E) ~= ~(a-b)\frac{2}{\pi}\lbrack\tan^{-1}(\frac{k\lambda(E)-3.4}{2.37})\rbrack + b
\end{equation}
This reduces the impact of the charged particle's energy on the induction of complex
damage in a DNA--molecule to three parameters $a$, $b$, and $k$. Figure \ref{keff}
illustrates the use of these parameters and shows that comparable results to the energy--based formalism
can be obtained. It follows that we can repeat the fitting procedure keeping $a$ and $b$ from the
expression based on energy (Eq \ref{response}). We find values of k=5.18 for
electrons and k=4.82 for protons. 
\subsection*{Extending the model} 
In the work presented above as well as in the used Monte Carlo
simulations, the Bethe--Barkas formalism  together with its flawed approach in
the lower energy regions has always been used. It is well established
that the IMFP does not follow the expression outlined in Equation
\ref{first_appr}, where $\lambda(E)$ keeps diminishing as the
energy diminishes. Indeed, when the energy is lower than 200eV an increase
in IMFP is observed due to plasmonic effects\cite{SIA:SIA1997}. Ziaja et al\cite{ziaja:033514} showed that it is
possible to describe this behavior analytically by extending Equation
\ref{first_appr} with a second term as follows:
\begin{equation}
\label{second_appr}
\lambda(E)_Z~=~ \frac{\sqrt{E}}{A_1(E-E_{th})^{B_1}} ~+~\frac{E - E_0\exp(-B/A)}{A\log(E/E_0)+B}
\end{equation}
In this equation the parameter $E_{th}$ serves as a threshold
separating the behavior as described by Bethe from the plasmonic
interactions. Using the data provided in the work from Zhen--Yu and
colleagues\cite{TANZhen-Yu:113403} it is straightforward to obtain
parameters for the behavior of protons. These are presented in Table
\ref{ziaja_table2}.
\begin{table}[h]
\centering
\begin{tabular} { | l | c | c | c | c | c | c |}
\hline
Particle&$A_1$&$B_1$&$E_{th}$&$A$&$B$&$E_0$\\\hline
Electrons&0.6560&1.0100&24.2838&65.898&-128.23&1.0\\\hline
Protons&0.681&1.249&42.38&117.01&-318.7&$1.0\times 10^3$\\\hline
\end{tabular}
\caption{\label{ziaja_table2} Parameters as in Table \ref{ziaja_table} with added lower energy factors. The fitting was performed 
using Eq. \ref{second_appr}}
\end{table}
\par
\begin{figure}[h]
\centering
\subfigure{\includegraphics[width=0.45\columnwidth]{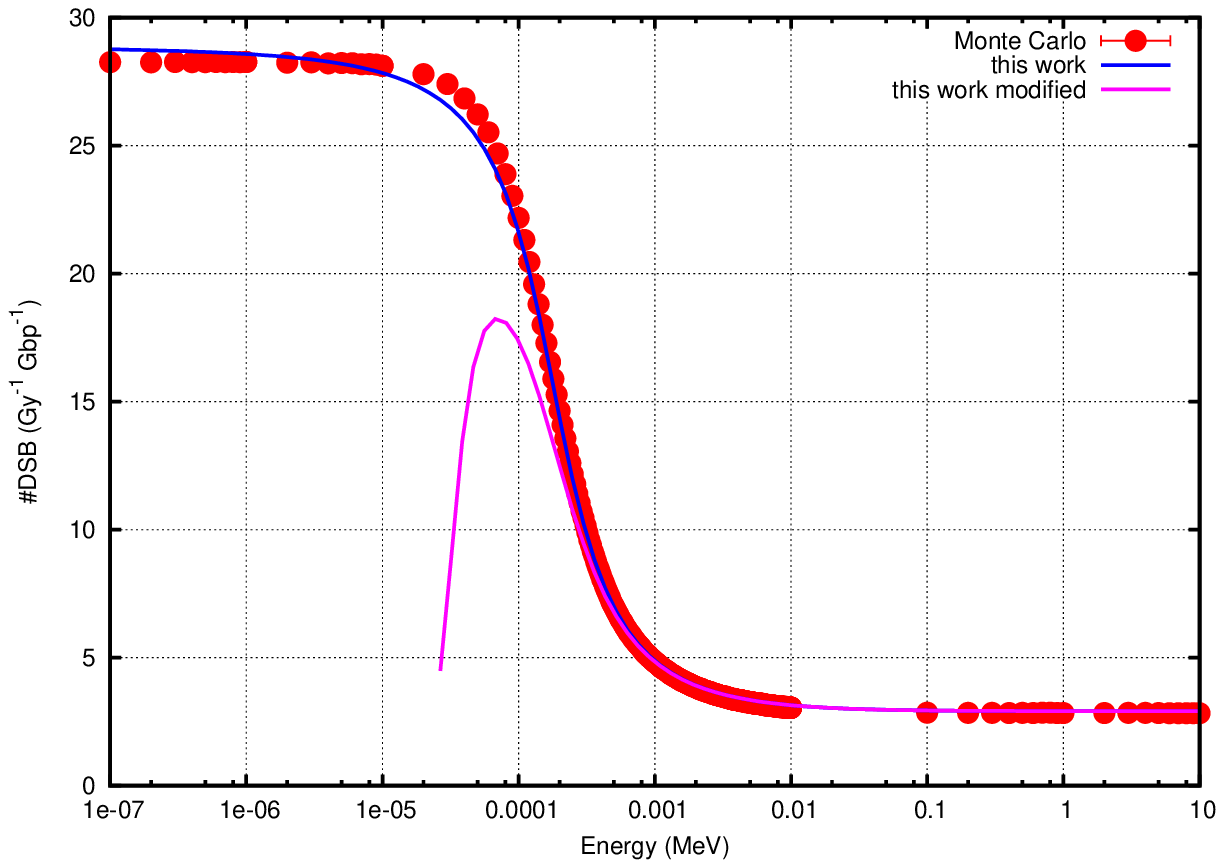}}
\subfigure{\includegraphics[width=0.45\columnwidth]{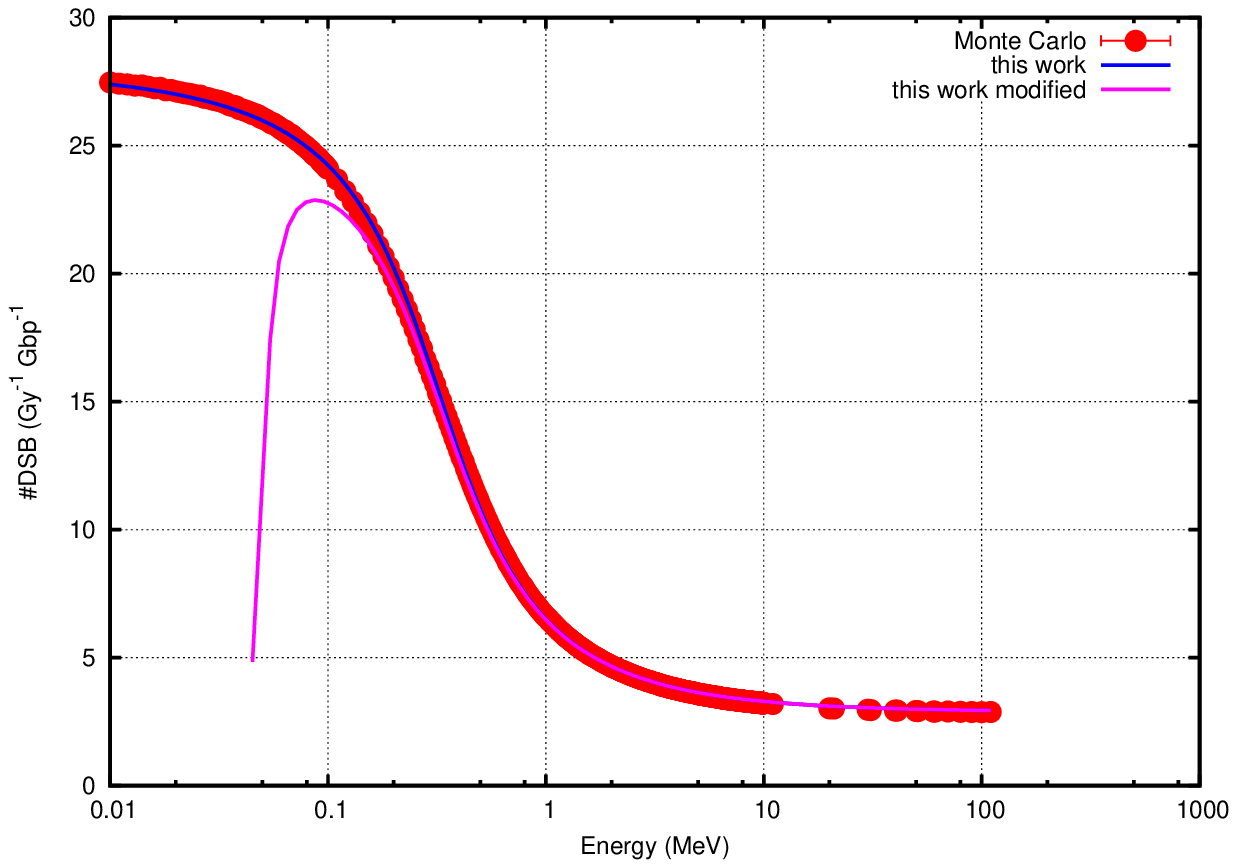}}
\caption{\label{keff} 
Using the quantities for $H$ and $r$, 
the dimensionless constant $k$ for electrons (left) and protons (right) is determined.
Using both the limited expression for $\lambda(E)$ and the more accurate estimate $\lambda_Z(E)$. The former 
provides a fit to the Monte Carlo data comparable with the results obtained using the energy--based formalism. 
The second approach provides a maximal complex damage yield which differs for electrons and protons.}
\end{figure}

To extend our model to incorporate the behavior of very low energy
particles it is sufficient to replace the expression $\lambda(E)$ by
$\lambda(E)_Z$ in equation \ref{Cauchy2}. In Figure \ref{keff}, the
modified curves show the difference with the calculations based on the
Bethe formalism only. This also shows that there is an upper limit to the
increase in DSB's which depends on the type of particle. It
is conceivable that this approach also works for the heavier particles
which  can be seen when using the IFMP's in water for these (not shown).
\section*{Discussion}
We developed an approach to predict damage in complicated situations
where fields of different charged particles and their
respective energy spectra impact
on living cells.
The approach, due to its analytical nature allows very fast calculation of damage in otherwise long simulations.
In the derivation of this approach using energy alone there are no
assumptions on the mechanics with which DNA--damage is caused by the
charged particles.
The only assumption is that there is a sensitive volume where, if
ionizations take place, damage is introduced in the DNA. How exactly
this damage is caused is not specified. In the remainder of the text a
parameter is identified, the average distance between ionizations
for the given charged particle in the medium ($\lambda$).  We show
that this approach adequately quantifies the results from Monte Carlo
simulations based on
phenomenological data and reduces these to a closed analytical expression
whereby the type of charged particle is expressed by a single parameter
($k$). 
On the other hand we should be aware that issues like
repair mechanisms and oxygen effects are not present in the model, making
its applicability limited. However, if all things are identical (i.e. the
type of cells, oxygenation, etc...) and the only thing different is
the type of charged particle and its energy, then the original damage
introduced in the DNA structure
should correlate with the outcome. An underlying assumption here is
that the repair processes are somehow independent from the modality with
which the cell is irradiated.
\par
The results of this approach can be applied to determine the
biological impact of radiation in mixed environments, as in the case
of proton therapy, where protons, electrons and heavier ions (due to
neutrons), deposit energy.
Other approaches have been proposed to try to predict outcomes from mixed
fields, which are based on available clinical response data. Most notably,
an approach based  on the local effect model (LEM), where macroscopic
response data in the form of dose--effect curves is used to quantify
the relative effect of the dose delivered\cite{Kraemer2006}. The
parameterization, however, of the latter approach is extensive due to
the fact
that every effect curve has two parameters for a given
$\alpha/\beta$--value, making the model over--parameterized. As such, it
is possible to
have this model reflect the current knowledge of dose and modality
response adequately, which forms an important, albeit controversial
tool\cite{Katz2003,doi:10.1080/09553000110066059}.
Its power to predict the behavior outside of the current knowledge therefore seems to be limited.
\par
Cucinotta et al. attempted to
incorporate the volumetric properties of the dose
deposition\cite{Cucinotta2000} to account for differences in track structure.
They observed that: ``LET is a poor
descriptor of energy deposition in small volumes  because of the
diffusion of secondary electrons out of the volume and contribution
of $\delta$--rays that pass outside of the volume''. To address this
problem a quantification of the energy distribution of generated secondary
particles, or $\delta$--rays was proposed.
\par
Such a secondary charged particle indirectly changes the behavior
with respect to the DNA damage induced. Indeed, depending on the median
energy of the spectrum the DNA damage changes accordingly if the dose
is kept constant. In the paper presented here this behavior could be
easily incorporated by considering the DNA damage for all the particles
(i.e. ions and $\delta$--rays) separately using a methodology modeled
on the use of the electronic equilibrium concept in photon cavity
theory. Currently this behavior is hidden in the $k$ parameter and it would be interesting to see if
such an approach will lead to a convergence of all $k$--values for all particles.
\par
To take these actions fully into account an approach to provide a more detailed model of the biological effect  directly in the Monte Carlo simulation is proposed by Sato et al.\cite{Sato2009} This would, in theory, allow a direct calculation of the effect in terms of energy deposited. However, as outlined by Cucinotta this is not without problems as the behavior of low energy electrons needs to be adequately modelled. This work predicts that the current knowledge using the Bethe formalism, might not be suitably extended.
\par

The results from the geometric interpretation indicate that the overall
behavior of the DNA damage induction is identical for all types of
charged particles. The only difference is in the dimensionless parameter
$k$. The latter seems to change as the ion used is heavier. Preliminary
calculations using the IMFP in water indicate that the value of $k$
diminishes as the
charged particles used are heavier (or more charged, data not shown). A possible
reason for this is that the track structure can be quite different for
different charged particles. This fact could also be an explanation for the discrepancy found at very low energies for carbon--ions. Indeed, allowing the parameter $k$ to be covariant with the other parameters, does provide a more adequate fit (data nor shown).
\par
The results for the electrons also shows a  discrepancy with regard
to the generation of complex damage at lower energies. For electrons, the data on very low energy electrons are not available in terms of energy deposition. Indeed, the model proposed here shows a much lower incidence of complex damage due to plasmonic effects in that region.
\par
In summary, the model proposed here allows extension
to very low energies for electrons and protons.The fact that there
are indications that the induction of DSB's varies
linearly with dose, provides an easy implementation to dose planning
systems, given the knowledge of dose deposition spectra in a treatment
beam. An example of such implementation is provided in the
Appendix.

\section*{Appendix: Implementation in dose deposition calculations}
\subsection*{Mono--energetic treatment}
In dose calculations a dose matrix is obtained on a dose grid
Let $\mathbf{D}=D[i,j,k]$ be the dose matrix provided.
Then we can
write the amount of complex damage incurred by particles with an energy ($E$)
as a damage matrix, denoted as ($M_{cd}$).
as follows:
\begin{align}
\mathbf{M_{cd}}= M_{cd}[i,j,k] &= \mathbf{D}\times  F_{cd}(E)
\end{align}
$F_{cd}(E)$ then denotes a response function converting dose to damage.
\subsection*{Poly--energetic treatment}
Dose deposition spectra rarely consist of a field of mono--energetic electrons.
For a photon source with a given photon spectrum, an energy depositing
electron fields exists, which is roughly constant throughout the target
volume. Using Monte Carlo simulations it is possible to calculate this field
and its spectrum $\Psi(E)$. It then becomes possible to include the
spectrum in the calculation of the damage matrices. This approach has been
used already by different authors \cite{Vandenheuvel-2010-nano,hsiao2008}.
\begin{equation}
\label{spectral_damage}
\mathbf{M_{cd}}~=~\mathbf{D}\times \frac{\int_0^{E_{max}}\Psi(E) F_{cd}(E)dE}{\int_0^{E_{max}}\Psi(E)dE}
\end{equation}
In the case of charged particle treatment, the particles are moderated
and the energy spectrum changes depending on the
position of the point where the dose is being deposited. It is therefore
necessary to apply Equation \ref{spectral_damage} to each
point separately with knowledge of the depositing energy spectrum in
that point. Due to the closed nature of the formalism developed in this paper, it becomes feasible to use off the shelf computing equipment.
\subsection*{Application: Proton treatment}
Recently, the coupling of Monte Carlo simulations in dose deposition to
micro-dosimetric code has been proposed and applied  by several groups\cite{hsiao2008,Vandenheuvel-2010-nano}.
Here a two step approach is followed; 1) a general purpose Monte Carlo code (MCNPX 2.7b)\cite{MCNPXref}
is used to estimate the spectrum of all different dose
contributing particles, 2) a micro dosimetric code\cite{ISI:000237044600004} is used to determine
the biological damage.
\par
\begin{figure}[h]
\centering
\subfigure[\label{protoncomb} Impact of the spectrum]{\includegraphics[width=0.45\columnwidth]{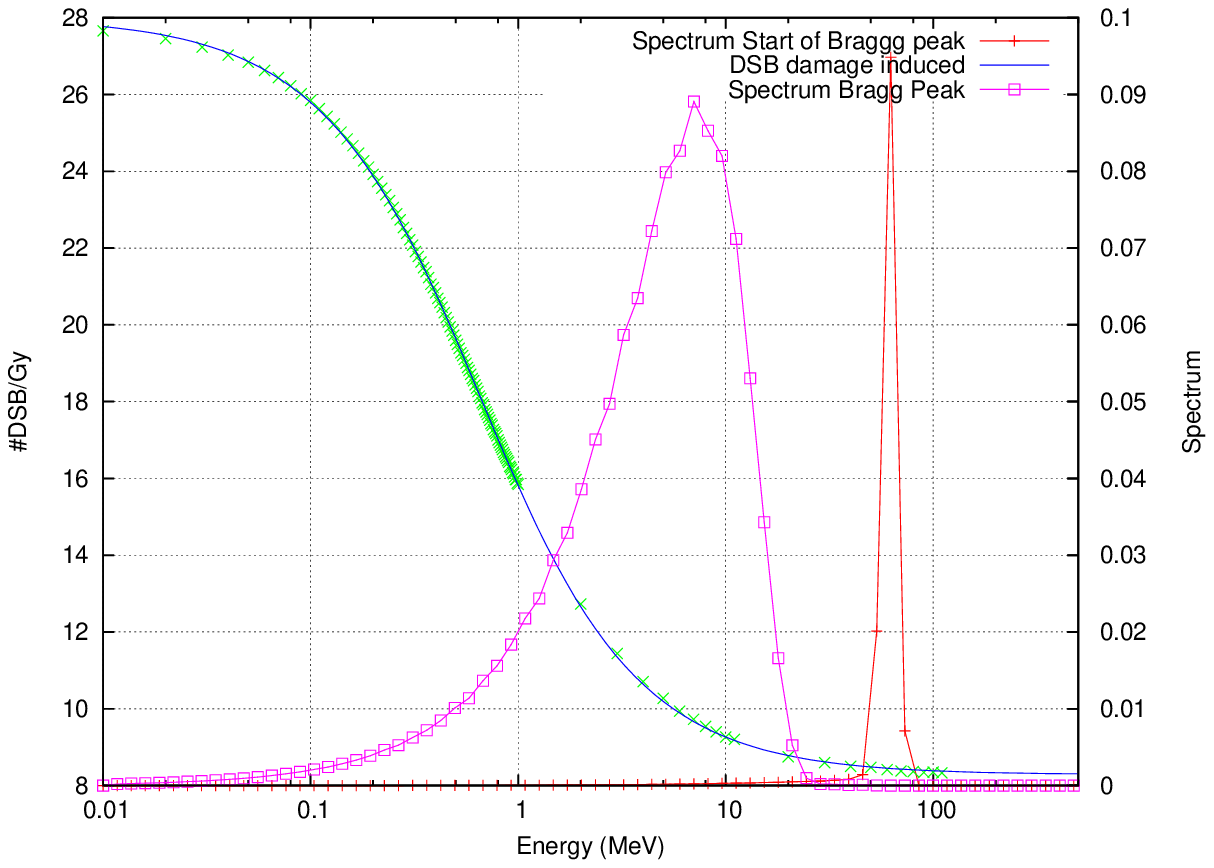}}
\subfigure[\label{peak} Proton dose deposition]{\includegraphics[width=0.45\columnwidth]{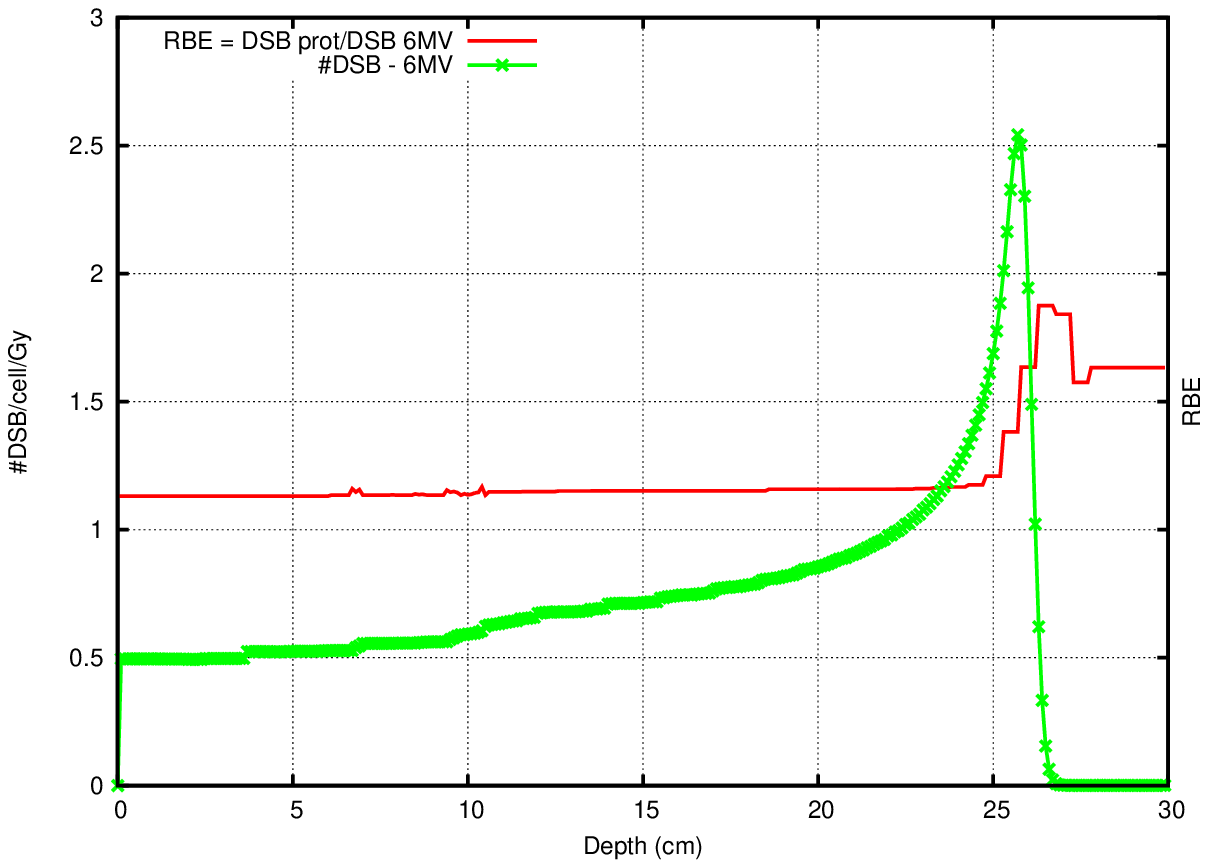}}
\caption{In Fig \ref{protoncomb}
the spectrum at the beginning of the Bragg peak (scaled by 0.1) is completely within the low LET
regimen. While at the end of the Bragg peak a significant part of the
dose depositing protons exhibits high LET characteristics.
Fig.\ref{peak} shows the $\mathrm{RBE_{cd}}$ (red line)  together with the damage induced
by a mono--energetic proton beam.}
\end{figure}

The framework for conversion of dose to biological effect is implemented
on a simulation of a pristine 200MeV proton beam, taking into account
the changing proton spectrum. The proton simulation is performed using
MCNPX. Figure \ref{protoncomb} shows the variation of the
number of complex damage events as a function of energy of the proton. In
addition, the spectrum of depositing protons is
shown at a position before the Bragg peak and at the Bragg peak.
In Figure \ref{peak} the effect on the dose deposition is shown together
with the $RBE_{cd}$ calculated as the complex damage yield generated by the protons
at that particular position, divided by the
complex damage induced by a 6MV photon beam with the same spatial characteristics.
Note, that the $RBE_{cd}$ is of the order of 1.1 with larger
value of 2 a few mm distal from the Bragg peak. This is commensurate
with the cell data reported by Paganetti et al.\cite{Paganetti2002} and Chaudhary et al.\cite{Chaudhary2014},
who showed that the radiobiological effect at the distal
end of a spread out bragg peak increases, a fact predicted by
Goitein\cite{goitein_book}. Currently, data of direct measurement
of DNA--damage in--vitro along a proton beam are scarce. The advent
of $\gamma$--H2AX measurements, as a marker for DSB--damage is promising
in this regard and has been used to investigate
anti--protons\cite{Kavanagh2013}.

\section*{Acknowledgements}
I am indebted to Rob Stewart, not only for generously providing the MCDS--code for everyone to use, but also for 
providing  much needed input on the biological aspects of the MCDS implementation. Both Ricardo Raabe and Mike Partridge reviewed some of the physics and made it a more rigorous paper. Sandra Nuyts and Dirk De Ruysscher's clinical input was also highly appreciated.


\begin{thebibliography}{10}
\providecommand{\url}[1]{\texttt{#1}}
\providecommand{\urlprefix}{URL }
\expandafter\ifx\csname urlstyle\endcsname\relax
  \providecommand{\doi}[1]{doi:\discretionary{}{}{}#1}\else
  \providecommand{\doi}{doi:\discretionary{}{}{}\begingroup
  \urlstyle{rm}\Url}\fi
\providecommand{\bibAnnoteFile}[1]{%
  \IfFileExists{#1}{\begin{quotation}\noindent\textsc{Key:} #1\\
  \textsc{Annotation:}\ \input{#1}\end{quotation}}{}}
\providecommand{\bibAnnote}[2]{%
  \begin{quotation}\noindent\textsc{Key:} #1\\
  \textsc{Annotation:}\ #2\end{quotation}}
\providecommand{\eprint}[2][]{\url{#2}}

\bibitem{Hall}
Hall EJ (1978) {LET} and {RBE}.
\newblock In: Radiobiology for the radiologist, Philadelphia: Harper \& Row,
  Publishers.
\input{Hall}

\bibitem{Caldecott2008}
Caldecott KW (2008) Single-strand break repair and genetic disease.
\newblock Nat Rev Genet 9: 619--631.
\input{Caldecott2008}

\bibitem{Ward1985}
Ward JF (1985) Biochemistry of dna lesions.
\newblock Radiation Research Supplement 8: pp. S103-S111.
\input{Ward1985}

\bibitem{Bethe30}
Bethe H (1930) Zur theorie des durchgangs schneller korpuskularstrahlen durch
  materie.
\newblock Ann Phys 5.
\input{Bethe30}

\bibitem{Barkas1963}
Barkas W, Dyer J, Heckman H (1963) Resolution of the $\sigma$--mass anomaly.
\newblock Phys Rev Lett 11: 26.
\input{Barkas1963}

\bibitem{doi:10.1080/09553009214551591}
Brenner DJ, Ward J (1992) Constraints on energy deposition and target size of
  multiply damaged sites associated with dna double-strand breaks.
\newblock International Journal of Radiation Biology 61: 737-748.
\input{doi:10.1080/09553009214551591}

\bibitem{icru16}
ICRU, {International Commission on Radiation Units and Measurements} (1970)
  Linear energy transfer.
\newblock ICRU Report 16.
\input{icru16}

\bibitem{icru19}
ICRU, {International Commission on Radiation Units and Measurements} (1971)
  Radiation quantities and units.
\newblock ICRU Report 19.
\input{icru19}

\bibitem{icru36}
ICRU, {International Commission on Radiation Units and Measurements} (1983)
  Microdosimetry.
\newblock ICRU Report 36.
\input{icru36}

\bibitem{Nikjoo1997}
Nikjoo H, O'Neill P, Goodhead DT, Terrissol M (1997) {Computational modelling
  of low-energy electron-induced DNA damage by early physical and chemical
  events}.
\newblock Int J Radiat Biol 71: 467-473.
\input{Nikjoo1997}

\bibitem{Stewart2011}
Stewart RD, Yu VK, Georgakilas AG, Koumenis C, Park JH, et~al. (2011) Effects
  of radiation quality and oxygen on clustered {DNA} lesions and cell death.
\newblock Radiation Research 176: 587-602.
\input{Stewart2011}

\bibitem{cauchy:1853}
Cauchy A (1853) Sur les r\'esultats moyens d'observations de m\^eme nature, et
  sur les r\'esultats les plus probables.
\newblock Comptes Rendus de l'Acad\'emie des Sciences : 198-206.
\input{cauchy:1853}

\bibitem{ISI:000237044600004}
Semenenko V, Stewart R ({2006}) {Fast Monte Carlo simulation of {DNA} damage
  formed by electrons and light ions}.
\newblock {Phys Med Biol} {51}: {1693-1706}.
\input{ISI:000237044600004}

\bibitem{Breit59}
Breit G (1959) Handbuch der Physik XLI/1.
\newblock Berlin, Heidelberg: Springer.
\input{Breit59}

\bibitem{NIST2010}
Powell C, Jablonski A (2010) Nist standard reference database 71.
\newblock In: {NIST} Electron Inelastic--Mean--Free--Path Database ---
  Version1.2, {NIST} Gaithersburg, MD.
\input{NIST2010}

\bibitem{TANZhen-Yu:113403}
Zhen-Yu T, Yue-Yuan X, Ming-Wen Z, Xiang-Dong L (2010) Proton inelastic mean
  free path in a group of organic materials in 0.05-10mev range.
\newblock Chinese Physics Letters 27: 113403.
\input{TANZhen-Yu:113403}

\bibitem{sinden1994dna}
Sinden R (1994) DNA structure and function.
\newblock San Diego: Academic Press.
\input{sinden1994dna}

\bibitem{SIA:SIA1997}
Tanuma S, Powell CJ, Penn DR (2005) Calculations of electron inelastic mean
  free paths.
\newblock Surface and Interface Analysis 37: 1--14.
\input{SIA:SIA1997}

\bibitem{ziaja:033514}
Ziaja B, London RA, Hajdu J (2006) Ionization by impact electrons in solids:
  Electron mean free path fitted over a wide energy range.
\newblock Journal of Applied Physics 99: 033514.
\input{ziaja:033514}

\bibitem{Kraemer2006}
Kraemer M, Scholz M (2006) Rapid calculation of biological effects in ion
  radiotherapy.
\newblock Phys Med Biol 51: 1959-1970.
\input{Kraemer2006}

\bibitem{Katz2003}
Katz R (2003) The parameter-free track structure model of {Scholz} and {Kraft}
  for heavy-ion cross sections.
\newblock Radiation Research 160: 724--728.
\input{Katz2003}

\bibitem{doi:10.1080/09553000110066059}
Paganetti H, Goitein M (2001) Biophysical modelling of proton radiation effects
  based on amorphous track models.
\newblock International Journal of Radiation Biology 77: 911-928.
\input{doi:10.1080/09553000110066059}

\bibitem{Cucinotta2000}
Cucinotta FA, Nikjoo H, Goodhead DT (2000) Model for radial dependence of
  frequency distributions for energy imparted in nanometer volumes from hze
  particles.
\newblock Radiation Research 153: 459--468.
\input{Cucinotta2000}

\bibitem{Sato2009}
Sato T, Kase Y, Watanabe R, Niita K, Sihver L (2009) Biological dose estimation
  for charged-particle therapy using an improved phits code coupled with a
  microdosimetric kinetic model.
\newblock Radiation Research 171: 107--117.
\input{Sato2009}

\bibitem{Vandenheuvel-2010-nano}
{Van den Heuvel} F, Locquet JP, Nuyts S (2010) Beam energy considerations for
  gold nano-particle enhanced radiation treatment.
\newblock Physics in Medicine and Biology 55: 4509.
\input{Vandenheuvel-2010-nano}

\bibitem{hsiao2008}
Hsiao Y, Stewart RD (2008) Monte carlo simulation of {DNA} damage induction by
  {X}-rays and selected radioisotopes.
\newblock Phys Med Biol 53: 233.
\input{hsiao2008}

\bibitem{MCNPXref}
Waters LS, McKinney GW, Durkee JW, Fensin ML, Hendricks JS, et~al. ({2007})
  {The MCNPX Monte Carlo radiation transport code}.
\newblock In: {Albrow, M and Raja, R}, editor, {Hadronic Shower Simulation
  Workshop}. {Amer Inst Physics}, volume {896} of \emph{{AIP conference
  proceedings}}, pp. {81-90}.
\newblock {Hadronic Shower Simulation Workshop, Batavia, IL, SEP 06-08, 2006}.
\input{MCNPXref}

\bibitem{Paganetti2002}
Paganetti H, Niemierko A, Ancukiewicz M, Gerweck LE, Goitein M, et~al. (2002)
  Relative biological effectiveness ({RBE}) values for proton beam therapy.
\newblock International journal of radiation oncology, biology, physics 53:
  407--421.
\input{Paganetti2002}

\bibitem{Chaudhary2014}
Chaudhary P, Marshall TI, Perozziello FM, Manti L, Currell FJ, et~al. (2014)
  Relative biological effectiveness variation along monoenergetic and modulated
  bragg peaks of a 62-mev therapeutic proton beam: A preclinical assessment.
\newblock Int J Radiat Oncol Biol Phys 90: 27--35.
\input{Chaudhary2014}

\bibitem{goitein_book}
Goitein M (2008) Radiation Oncology: A Physicist's-Eye View.
\newblock Springer Verlag.
\input{goitein_book}

\bibitem{Kavanagh2013}
Kavanagh JN, Currell FJ, Timson DJ, Savage KI, Richard DJ, et~al. (2013)
  Antiproton induced dna damage: proton like in flight, carbon-ion like near
  rest.
\newblock Sci Rep 3: --.
\input{Kavanagh2013}

\end{thebibliography}

\end{document}